\documentclass[11pt]{article}
\usepackage{amssymb,amsmath,amsfonts}
\usepackage{graphicx}
\usepackage{graphics}
\usepackage{eepic,epsfig}
\usepackage{cite}

\begin{document}


\title{The Zeta Function Approach for Casimir Energy Calculations in Higher Dimensions}
\author{R. K. Obousy\thanks{%
E-mail: robousy@icarusinterstellar.org } \\
\textit{Casimir Energy Research Group, Icarus Interstellar Inc., }\\
\textit{Longview, Texas, USA, 75604 }}

\maketitle

\section{Abstract}

The vacuum fluctuations give rise to a number of phenomena; however, the the Casimir Effect is arguably the most salient manifestation of the quantum vacuum. In its most basic form it is realized through the interaction of a pair of neutral parallel conducting plates. The presence of the plates modifies the quantum vacuum, and this modifcation causes the plates to be pulled toward each other. The Casimir Effect has also been explored in the context of higher dimensional theories. The non-trivial boundary conditions imposed by compactified periodic higher dimensions is know to alter the vacuum in a quantifiable way, and is a possible solution to the issue of modulus stabilization, namely the stabilization of higher dimensions. 

Typical in Casimir energy calculations are renormalization techniques which are used to tame the infinite sums and integrals that arise. These calculations are usally fairly involved, and $\textit{explicit}$ pedagogical material is sparse. The purpose of this paper is to introduce $\zeta$-function regularization techniques specific to Casimir energy calculations in an $M_4 \times T^1$ spacetime.

\section{Introduction}

Compactified extra dimensions introduce non-trivial boundary conditions to the quantum vacuum and Casimir-type calculations become important when finding the resulting vacuum energy. Although tantalizing, the study of extra dimensions opens up a whole new set of questions. For example, if extra dimensions exist and are hidden from us due to their compact nature, then what \textit{keeps} them small? Why do they not expand to be large like the three macroscopic dimensions we are familiar with, or conversely, why not perpetually shrink? 

One possibile answer is that the Casimir energy spectrum generates a stable minimum of the higher dimensional vacuum. This is covered more explicitly in \cite{obousy}. Before we can explore such interesting scenerios we must first learn the mathematical machinery that will allow us to examine this further. We begin, in Section 3, by introducing $\zeta$ function regularization techniques for the case of a one dimensional mnassless scalar field. Then, in Section 4, we will expand the dimensionality of the problem up to an $M_4 \times T^1$ spacetime. Finally, in Section 5 we will study a massive scalar field with both periodic and non-periodic boundary conditions in the full $M_4 \times T^1$ spacetime. For further reading on this fascinating field the reader is directed to \cite{Most97,ac, ac1, rr,I,Blau84, Cand84, Gilb84, Kikk85, Acce86, Maed87, Gunt97, nos, Gold00, Garr01, Pont01, Flac01, m, fmt, hkp, kt, nos1, enoo, gpt1, Saha04a, Saha05}.

\section{One Dimensional Massless Scalar Field Calculation}

Typically, the calculations of the expectation value of the vacuum is divergent, so some form of renormalization must be performed. For example, consider the calculation of the field vacuum expectation value (VEV) inside a metal cavity. Such a calculation will necessarily involve summing the energies of the standing waves in the cavity;
\begin{equation}
\left<E_{\rm vac}\right>=\frac{1}{2}\sum_{n=1}^\infty E_{\rm n},
\end{equation}
which is clearly divergent.

To calculate a finite value we will use a $\zeta$ function renormalization technique. Starting from basic quantum mechanics we know:

\begin{equation}
E=c\bf{p},
\end{equation}

and

\begin{equation}
p=\hbar \bf{k},
\end{equation}

and so

\begin{equation}
E=\hbar c \bf{k},
\end{equation}

For the remainder of this paper we will set $c$ and $\hbar$ to zero.

We first consider a 1D cavity of lenth $L$ bounded by perfectly conducting walls. Our boundary conditions imply that

\begin{equation}
\rm{sin}(kL)=\rm{sin}(n\pi),
\end{equation}

and so we can quickly obtain:

\begin{equation}
k= \frac{n \pi}{L}. 
\end{equation}
For every k there exists two standing waves unless $n$ is zero. To calculate the energy between the plates we need to some over all possible modes.
\begin{equation}
E_0(L)=\frac{\pi}{2L}\sum_{n=1}^\infty n, \ 
\label{eq87}
\end{equation}
that is, we sum over all modes `n' of the vacuum.
By utilizing the Riemman $\zeta$ function we can deftly handle this infinite sum.
\begin{equation}
\zeta(s)=\sum_{n=1}^\infty \frac{1}{n^s},
\label{eq88}
\end{equation}
we can rewrite eq.\ (\ref{eq87}) as:
\begin{equation}
E_0(L)=\frac{\pi}{2L}\sum_{n=1}^\infty \frac{1}{n^{-1}}=\frac{\pi}{2L}\zeta(-1). \
\label{eq89}
\end{equation}
However, $\zeta(-1)=-\frac{1}{12}$ (from analytic continuation), and so we quickly obtain the result:
\begin{equation}
E_0(L)=-\frac{\pi}{24L},
\label{eq90}
\end{equation}
which is the Casimir energy. Note that one of the salient features of Casimir energy is that it is \textit{negative}. Taking the derivative, we obtain the force of attraction between the plates:
\begin{equation}
\mathcal{F}(L)=-\frac{\partial}{\partial L}E_0(L)=\frac{\pi}{24L^2}.
\label{eq91}
\end{equation}	
Note that this is the Casimir energy of a 1D massless scalar field, so differs from the more familiar $\frac{1}{L^4}$ relationship that some readers may be familiar with.

\section{$M_4 \times T^1$ Massless Scalar Field Calculation}

In the presence of an additional compact circular dimension, the expression for the energy of a massless scalar field becomes

\begin{equation}
E^2=k^2+ \left( \frac{n\pi}{R} \right) ^2,
\label{eq125}
\end{equation}

where the extra term is due to the compactification. We now calculate the higher dimensional Casimir energy density, recalling that:

\begin{equation}
\frac{1}{V}\sum_k\rightarrow\frac{1}{(2\pi)^3}\int d^3k,
\end{equation}

which is modified for a 4+1 dimensional space to 

\begin{equation}
\frac{1}{V^{(4)}}\sum_k\rightarrow\frac{1}{(2\pi)^4}\int d^4k.
\end{equation}

The Casimir energy density due to the KK modes of a scalar field, obeying periodic boundary conditions compactified on $S^1$, is  
\begin{equation}
E=\frac{1}{2}{\sum_{n=-\infty}^{\infty}}'\int\frac{d^4k}{(2\pi)^4}ln\left( k^2+\left(\frac{n\pi}{R} \right)^2         \right), 
\label{eq126}
\end{equation}
where the prime on the summation indicates that the $m=0$ term is omitted, as it represents a constant energy which we can subtract. The log expression for the energy is a popular starting point for regularization schemes as it allows for the use of the Gamma function. It's thermodynamically equivalent to the free energy at zero temperature (see for example Finite Temperature Field Theory, Kapusta, Ch2)

We can rewrite the log as a derivative, perform a dimensional regularization on the integral and the summation
\begin{eqnarray}
E &=&\frac{1}{2}\frac{\partial}{\partial s}|_{s=0}{\sum_{n=-\infty}^{\infty}}'\int\frac{d^4k}{(2\pi)^4}\left( k^2+\xi n^2 \right)^{-s} \nonumber \\
  &=&\frac{1}{2}\frac{\partial}{\partial s}\zeta^+(s) |_{s=0}
  \label{eq127},
\end{eqnarray}
where the periodic scalar function is defined as
\begin{equation}
\zeta^+(s)={\sum_{n=-\infty}^{\infty}}'\int\frac{d^4k}{(2\pi)^4}\frac{1}{\Gamma(s)}\int_0^\infty dte^{(k^2+\xi n^2)t}t^{s-1}. 
\label{eq129}
\end{equation}
Here we have made the substitution $\xi=\frac{\pi^2}{R^2}$ and used the identity
\begin{equation}
z^{-s}=\frac{1}{\Gamma(s)}\int_0^\infty dt e^{-zt}t^{s-1}.
\label{eq130}
\end{equation}
We first perform the k integral, 
\begin{equation}
\int d^4k e^{-k^2t}=\frac{\pi^2}{16t^2},
\label{eq131}
\end{equation}
and now calculate
\begin{equation}
\zeta^+(s)=\frac{\pi^2}{(2\pi)^4}\frac{1}{\Gamma(s)}{\sum_{n=-\infty}^{\infty}}'\int_0^\infty dt e^{-\xi n^2t}t^{s-3}.
\label{eq132}
\end{equation}
Making the substitution $x=\xi n^2 t$ gives us
\begin{eqnarray}
t=\frac{x}{\xi n^2}\quad{\rm and}\quad dt=\frac{dx}{\xi n^2}.
\label{eq133}\end{eqnarray}
Now substituting back into eq.\ (\ref{eq132}),
\begin{equation}
\zeta^+(s)=\frac{\pi^2}{(2\pi)^4}\frac{1}{\Gamma(s)}{\sum_{n=-\infty}^{\infty}}'\int_0^\infty \frac{dx}{\xi n^2} e^{-x}\left(\frac{x}{\xi n^2}\right)^{s-3}
\label{eq134}.
\end{equation}
We can express the $x$ integral in terms of the Gamma function
\begin{equation}
\zeta^+(s)=\frac{\xi^{2-s} \pi^2}{(2\pi)^4}\frac{\Gamma(s-2)}{\Gamma(s)}   {\sum_{n=-\infty}^{\infty}}'\frac{1}{n^{2s-4}}.
\label{eq135}
\end{equation}
We immediately recognise the infinite sum as the Riemann Zeta function, so we finally obtain
\begin{equation}
\zeta^+(s)=\frac{\xi^{2-s} \pi^2}{(2\pi)^4}\frac{\Gamma(s-2)}{\Gamma(s)}   \zeta(2s-4).
\label{eq135a}
\end{equation}
Using the recursion relationship for the Gamma function, as defined in this appendix, we can express this function as 
\begin{equation}
\frac{\Gamma(s-2)}{\Gamma(s)}=\frac{\Gamma(s-2)}{(s-2)(s-1)\Gamma(s-2)}, 
\label{eq136}
\end{equation}
Now, plugging this expression back into eq.\ (\ref{eq127}), and performing the derivative with respect to s evaluated at $s=0$ we obtain
\begin{equation}
E=-\frac{\pi^2}{2\pi^4} \left(    \frac{\pi^2}{R^2}\right)^2\zeta'(-4).
\label{eq137}
\end{equation}
However, the derivative of the zeta function is known to be
\begin{equation}
\zeta'(-4)=\frac{3}{4\pi^4}\zeta(5),
\label{eq138}
\end{equation}    
and so we find our final expression for the Casmir energy density of a scalar field with periodic boundary conditions to be
\begin{equation}
E=-\frac{3}{64\pi^2}\frac{1}{R^4}\zeta(5).
\label{eq139}
\end{equation}
We see that the Casimir energy density is proportional to $1/R^4$, where $R$ is the size of the extra dimension.

\newpage

\section{$M^4 \times T^1$ Massive Scalar Field - Periodic and Non-Periodic Fields}

The Casimir energy density generated from the quantum fluctuations in the large dimensions are insignificant when they are compared to the contributions arising from the compact dimensions, because the energy is inversely proportional to volume of the space. Therefore, this Casimir energy calculation focuses on the Casimir energy for a field with boundary conditions on the $T^1$ compactification (a circle). We will again use $\zeta$-function techniques. 

For a massive field, we can express the modes of the vacuum as 
\begin{equation}
E_n=\sqrt{{\bf k}^2+\left(\frac{\pi n}{r_c}\right)^2+M_n^2},
\label{eq147}
\end{equation}
with $M_n$ is the mass of the fields, and $r_c$ the radius of the compact extra dimension. As usual, we have used natural units. The Casimir energy is given by
\begin{equation}
V^+=\frac{1}{2}{\sum_{n=-\infty}^{\infty}}' \int \frac{d^4k}{(2\pi)^4}{\rm ln}({\bf k}^2+\left(\frac{n\pi}{r_c}\right)^2+M_n^2),
\label{eq148}
\end{equation}
where the prime on the summation indicates that the $n=0$ term is excluded. For purposes of regularization, we will write this as
\begin{equation}
V^+=\frac{1}{2}{\sum_{n=-\infty}^{\infty}}'\int\frac{d^4k}{(2\pi)^4}\int_0^\infty\frac{ds}{s}e^{-({\bf k}^2+\left(\frac{n\pi}{r_c}\right)^2+M_n^2)s}.
\label{eq149}
\end{equation}
We first perform the Gaussian integration (the k-integral)
\begin{equation}
\int_0^\infty d^4ke^{-{\bf k}^2s}=\frac{\pi^2}{s^2},
\label{eq150}
\end{equation}
and are left with the remaining calculation;
\begin{equation}
V^+=\frac{1}{2}\frac{\pi^2}{(2\pi)^4} {\sum_{n=-\infty}^{\infty}}'\int_0^\infty ds \ \frac{1}{s^3}e^{-\left((\frac{n\pi}{r_c})^2+M_n^2\right)s}.
\label{eq151}
\end{equation}
To help us solve this equation we will use the Poisson Resummation formula:\footnote{Sometimes called Jacobi's theta function identity.}
\begin{equation}
{\sum_{n=-\infty}^\infty}' e^{-(n+z)^2t}=\sqrt{\frac{\pi}{t}}\sum_{n=1}^\infty e^{-\pi^2n^2/t}cos(2\pi nz),
\label{eq152}
\end{equation}
to rewrite the summation of eq.\ (\ref{eq151}). Setting $z=0$ we obtain,
\begin{equation}
\sum_{n=-\infty}^\infty e^{-(\frac{n\pi}{r_c})^2}=\sqrt{\frac{1}{\pi s}}\sum_{n=1}^\infty e^{r_c^2n^2/s}.
\label{eq153}
\end{equation}
Inserting this back into eq.\ (\ref{eq151}) we see that our exponential term can now be expressed as
\begin{equation}
\frac{1}{2}\frac{\pi^2}{(2\pi)^4} \sum_{n=1}^\infty \sqrt{\frac{1}{\pi s}}e^{-(r_c^2n^2/s+M_n^2s)},
\label{eq154}
\end{equation} 
and inserting back into eq.\ (\ref{eq151}) our expression for the Casimir energy density now becomes
\begin{equation}
V^+=\frac{1}{2}\frac{\pi^2}{(2\pi)^4} \sqrt{\frac{1}{\pi}}\sum_{n=1}^\infty \int_0^\infty ds \ . \frac{1}{s^{7/2}}e^{-(M_nr_cn(\frac{M_ns}{r_cn}+\frac{r_cn}{M_ns}))}.
\label{eq155}
\end{equation} 
If we now set $x=\frac{M_ns}{r_cn}$ we can write our equation as:
\begin{equation}
V^+=\frac{1}{2}\frac{\pi^2}{(2\pi)^4} r_c^{-5/2}M_n^{5/2}\sum_{n=1}^\infty \frac{1}{n^{5/2}}\int_0^\infty dx x^{-7/2}e^{-M_nr_cn(x+\frac{1}{x})}.
\label{eq156}
\end{equation} 
The integral is easily solved using the following expression for the Modified Bessel function of the Second kind:
\begin{equation}
K_\nu(z)=\frac{1}{2}\int_0^\infty dx x^{\nu-1}e^{-z/2(x+\frac{1}{x})}.
\label{eq157}
\end{equation} 
Using eq. (39) in eq.\ (\ref{eq156}), and recognizing the infinite sum as the Riemann zeta function, we obtain our final expression for the Casimir energy density of a massive scalar field in the five dimensional setup.
\begin{equation}
V^+=-\frac{\zeta(5/2)}{32\pi^2}\frac{M_n^{5/2}}{r_c^{5/2}}\sum_{n=1}^\infty K_{5/2}(2M_nr_cn).
\label{eq158}
\end{equation} 
It is straightforward to extend this expression to include antiperiodic fields. Recalling eq.\ (\ref{eq152}), we see that for antiperiodic fields we can make the substition $n \rightarrow n+1/2$ which ensures the summation is over integer multiples of 1/2. This implies our z term in the Poisson Resummation fomula is now non-zero $(z=1/2)$, so we simply have to include the cosine term in our final Casimir energy expression. Thus, the Casimir energy for anti-periodic fields in our five dimensional setup becomes
\begin{equation}
V^-=-\frac{\zeta(5/2)}{32\pi^2}\frac{M_n^{5/2}}{r_c^{5/2}}\sum_{n=1}^\infty K_{5/2}(2M_nr_cn){\rm cos}(n \pi).
\label{eq159}
\end{equation}
We now wish to find an expression of the Casimir energy due to a \textit{massless} scalar, which will also be used as a component in the stabilization investigation. The necessary calculation is
\begin{equation}
V^+_{\rm massless}=\frac{1}{2}{\sum_{n=-\infty}^{\infty}}'\int\frac{d^4k}{(2\pi)^4}ln(k^2+\left(\frac{n\pi}{r_c}\right)^2)
\label{eq160}
\end{equation}
This calculation is well-known in the literature, and so we simply quote the result
\begin{equation}
V^+_{\rm massless}=-\frac{3\zeta(5)}{64\pi^2}\frac{1}{r_c^4}.
\label{eq161}
\end{equation}

From the expression for the Casimir contribution for a periodic massive scalar field it is straightforward to enumerate the Casimir contributions of all other massive and massless fields by using knowledge of five-dimensional supersymmetry multiplets \cite{Pont01},
\begin{equation}
V^+_{\rm fermion}(r)=-4V^{+}(r),
\label{eq23a}
\end{equation}
\begin{equation}
V^-_{\rm fermion}(r)=\frac{15}{4}V^{+}(r),
\label{eq23b}
\end{equation}
\begin{equation}
V^+_{\rm higgs}(r)=2V^{+}(r),
\label{eq23c}
\end{equation}
where the positive sign on the potential indicated a periodic field and a negative sign indicates an antiperiodic field.

\section{Acknowledgements}

I would like to acknowledge the Institute for Advanced Studies at Austin, and Eric Davis for motivating this paper.

\section{Appendix}

This section contains some very useful mathematical identities that are used in this paper.

\subsection{The Riemann Zeta Function}

\begin{equation}
\zeta(s)=\sum_{n=1}^\infty \frac{1}{n^s}
\end{equation}

*Note $\zeta(-1)=-\frac{1}{12}$

\subsection{Gaussian integration}

\begin{equation}
\int_0^\infty e^{-ax^2}dx=\frac{1}{2}\sqrt{\frac{\pi}{a}}
\end{equation}

Generalized to n dimensions:

\begin{equation}
\int_0^\infty e^{-ax^2}d^nx=\frac{1}{2^n} \left(\frac{\pi}{a}\right)^{n/2}
\end{equation}

\subsection{Tricks with Logs}

We can express a log as: 

\begin{equation}
ln(x)=-\frac{d}{ds}(x^{-s})|_{s=0}
\end{equation}

This is because:

\begin{equation}
x^{-s}=e^{ln(x^{-s})}=e^{-slnx}
\end{equation}

taking the derivative:

\begin{equation}
\frac{d}{ds}x^{-s}|_{s=0}=\frac{d}{ds}e^{-slnx}|_{s=0}=ln(x)e^0=ln(x)
\end{equation}

\subsection{The Gamma Function}

The Gamma function is given by:

\begin{equation}
\Gamma(n)=\int_0^\infty t^{n-1}e^{-t}dt
\end{equation}

For $n>1$. See, for example, Schaums p146, 25.1.

An alternative expression for the Gamma fucntion is:

\begin{equation}
\Gamma(n)=z^n\int_0^\infty t^{n-1}e^{-zt}dt
\end{equation}

The recursion formula for the Gamma function is given by

\begin{equation}
n\Gamma(n)=\Gamma(n+1)
\end{equation}

See, for example, Schaums p146, 25.2. Tip - to solve in example above, try using this formula twice, first for $n=(s-1)$, then for $n=(s-2)$.

\subsection{Poisson Resummation formula}

\begin{equation}
{\sum_{n=-\infty}^\infty}' e^{-(n+z)^2t}=\sqrt{\frac{\pi}{t}}\sum_{n=1}^\infty e^{-\pi^2n^2/t}cos(2\pi nz),
\end{equation}

\subsection{Modified Bessel Function of the Second Kind}

\begin{equation}
K_\nu(z)=\frac{1}{2}\int_0^\infty dx x^{\nu-1}e^{-(z/2)(x+\frac{1}{x})}.
\label{eq157}
\end{equation}

\end{document}